\documentclass[fleqn,usenatbib,letters]{mnras}
\usepackage[T1]{fontenc}
\usepackage{ae,aecompl}
\usepackage{pdflscape}

\usepackage{graphicx}	
\usepackage{amsmath}	
\usepackage{amssymb}	
\usepackage{bm}
\usepackage{mathtools}
\usepackage{float}
\usepackage{array, makecell}
\usepackage{booktabs}
\usepackage{hyperref}
\usepackage[svgnames]{xcolor}
\usepackage{xspace}
\usepackage{multirow}
\usepackage{newtxtext,newtxmath}

\renewcommand{\vec}[1]{{\bm{#1}}}
\newcommand{\oper}[1]{{\bm{\mathsf{#1}}}}
\newcommand{\software}[1]{{\textsc{#1}}}
\newcommand{\dft}[0]{\oper{D}}
\newcommand{\lensop}[0]{\oper{L}}
\newcommand{\source}[0]{\vec{s}}
\newcommand{\lams}[0]{\lambda_{\source}}
\newcommand{\data}[0]{\vec{d}}
\newcommand{\etalens}[0]{\vec{\eta}}
\newcommand{\model}[0]{\vec{m}}
\newcommand{\dpsi}[0]{\delta\vec{\psi}}
\newcommand{\post}[0]{\mathcal{P}}
\newcommand{\psmooth}[0]{\mathcal{P}_\mathrm{smooth}}
\newcommand{\prmean}[0]{\vec{\mu}_{\etalens, \lams}}
\newcommand{\prcov}[0]{\vec{\Sigma}_{\etalens, \lams}}
\newcommand{\msun}[0]{M_\odot}
\newcommand{\mdm}[0]{m_\chi}
\newcommand{\mdmc}[1]{m_{\chi,#1}}
\newcommand{\ldb}[0]{\lambdabar\chi}
\newcommand{\fdm}[0]{f_\mathrm{DM}}
\newcommand{\sigv}[0]{\sigma_v}
\newcommand{\pthresh}[0]{P_\mathrm{thresh}}
\newcommand{\units}[1]{~\mathrm{#1}}

\newcommand{\figscale}{0.75}

\newcommand*{\email}[1]{\href{mailto:#1}{\texttt{#1}}}
\newcommand{\jzsfo}[0]{MG J0751+2716\xspace}
\newcommand{\sersic}[0]{S\'{e}rsic\xspace}

\title[Fuzzy dark matter and \jzsfo]{A lensed radio jet at milli-arcsecond resolution II: Constraints on fuzzy dark matter from an extended gravitational arc}

\author[D. Powell et al.]{%
Devon M. Powell,$^{1}$\thanks{E-mail: \email{dmpowell@mpa-garching.mpg.de}}
Simona Vegetti,$^{1}$
J. P. McKean,$^{2,3}$
Simon D.M. White,$^{1}$
\newauthor
Elisa G. M. Ferreira,$^{4,5}$
Simon May$^{6,7}$,
and Cristiana Spingola$^{8}$
\\
$^{1}$Max Planck Institute for Astrophysics, Karl-Schwarzschild-Stra\ss{}e 1, 85748 Garching bei M\"unchen, Germany\\
$^{2}$Kapteyn Astronomical Institute, University of Groningen, PO Box 800, NL-9700 AV Groningen, The Netherlands\\
$^{3}$ASTRON, Netherlands Institute for Radio Astronomy, PO Box 2, NL-7990 AA Dwingeloo, The Netherlands\\
$^{4}$Kavli Institute for the Physics and Mathematics of the Universe (WPI), UTIAS, The University of Tokyo, Chiba 277-8583, Japan \\
$^{5}$Instituto de F\'{i}sica, Universidade de S\~{a}o Paulo, Rua do Mat\~{a}o 1371, Butant\~{a}, 05508-090, S\~{a}o Paulo, Brazil
\\
$^{6}$Perimeter Institute for Theoretical Physics, 31 Caroline Street North, Waterloo, ON, N2L 2Y5, Canada\\
$^{7}$Department of Physics, North Carolina State University, Raleigh, NC, 27695-8202, USA\\
$^{8}$INAF $-$ Istituto di Radioastronomia, via Gobetti 101, I$-$40129, Bologna, Italy\\
\\
}

\date{Accepted 2023 June 06. Received 2023 May 12; in original form 2023 February 21}

\pubyear{2023}

\begin{document}
\label{firstpage}
\pagerange{\pageref{firstpage}--\pageref{lastpage}} 
\maketitle

\begin{abstract}
  Using a single gravitational lens system observed at $\lesssim5$ milli-arcsecond resolution with very long baseline interferometry (VLBI), we place a lower bound on the mass of the fuzzy dark matter (FDM) particle, ruling out $\mdm \leq 4.4\times10^{-21}\units{eV}$ with a 20:1 posterior odds ratio relative to a smooth lens model. We generalize our result to non-scalar and multiple-field models, such as vector FDM, with $\mdmc{\mathrm{vec}} > 1.4 \times 10^{-21}\units{eV}$. Due to the extended source structure and high angular resolution of the observation, our analysis is directly sensitive to the presence of granule structures in the main dark matter halo of the lens, which is the most generic prediction of FDM theories. A model based on well-understood physics of ultra-light dark matter fields in a gravitational potential well makes our result robust to a wide range of assumed dark matter fractions and velocity dispersions in the lens galaxy. Our result is competitive with other lower bounds on $\mdm$ from past analyses, which rely on intermediate modelling of structure formation and/or baryonic effects. Higher resolution observations taken at 10 to 100 GHz could improve our constraints by up to 2 orders of magnitude in the future.
\end{abstract}

\begin{keywords}
gravitational  lensing:  strong --  galaxies: haloes -- cosmology: dark matter -- radio  continuum:  general -- quasars: individual: MG J0751+2716
\end{keywords}


\section{Introduction} \label{sec:intro}

The characterization of dark matter (DM) is of central importance to astrophysics and cosmology. Despite abundant observational evidence of a dark, collisionless fluid comprising $\sim 85$ per cent of the matter in the universe, the nature of dark matter remains an open question. While cold dark matter (CDM) is the current theoretical paradigm due to its success in explaining observed phenomena across a wide range of physical scales, evidence for its agreement with observations on sub-galactic scales has not been conclusive \citep[e.\,g.][]{bullock2017}.  Alternative models of dark matter comprised of ultra-light particles (ULDM) have been proposed as a way of alleviating such discrepancies without invoking complex baryonic feedback processes (see \citealt{ferreira2021} for a comprehensive review). 

Fuzzy Dark Matter (FDM) is a class of ULDM comprised of non-interacting scalar particles of mass $\mdm \sim 10^{-22}\units{eV}$. Due to its $\sim$kpc-scale de Broglie wavelength, FDM exhibits a rich astrophysical phenomenology \citep{hu2000,hui2017}. A key prediction from FDM models is that the mass density profiles of dark matter haloes exhibit small-scale fluctuations due to wave interference  (commonly termed ``granules''), which give them a vastly different structure from the haloes expected in the CDM and warm dark matter (WDM) models \citep{schive2014b,schive2016,may2022}.  In addition to these effects, on sub-galactic scales, FDM models predict much lower concentrations and cored density profiles in dwarf galaxies due to ``quantum pressure,'' which is absent in CDM and WDM \citep{schive2016}. Similarly to WDM, FDM also predicts a suppression in the numbers of low-mass haloes relative to CDM \citep{schive2016}, albeit via a different mechanism.

Constraints have been placed on the allowed mass range for the FDM particle via several observational routes.  Jeans modelling of dwarf spheroidal galaxies (dSphs) yields a lower bound of $\mdm \gtrsim 10^{-22}\units{eV}$ \citep{chen2017,saf2020,Hayashi:2021xxu}.  \citet{dalal2022} find $\mdm > 3\times10^{-19}\units{eV}$ by considering stellar velocity dispersions in ultra-faint dwarf galaxies (UFDGs).  Using Lyman-$\alpha$ forest observations of cosmic structure, \citet{irsic2017} and subsequently \citet{rogers2021} constrain $\mdm > 3.8\times10^{-21}\units{eV}$ and $\mdm > 2\times10^{-20}\units{eV}$, respectively. Constraints based on the Milky Way subhalo population include $\mdm > 2.9\times10^{-21}\units{eV}$ from number statistics of the observed Milky Way satellites \citep{nadler2021a} and $\mdm > 2.2\times10^{-21}\units{eV}$ using stellar streams in the Milky Way \citep{banik2021}.  A study of flux ratio anomalies in 11 quadruply-imaged gravitationally lensed quasars gives a lower bound of $\mdm > 10^{-21}$ eV \citep{laroche2022}.

In this work, we study a single observation of the gravitationally lensed radio jet \jzsfo. These data were taken at 1.6 GHz using global very long baseline interferometry (VLBI) with an angular resolution, measured as the full width at half maximum (FWHM) of the main lobe of the dirty beam response, of $5.5 \times 1.8 \units{mas}^2$. The details of this dataset were previously reported by \citet{spingola2018} and \citet{powell2022}.  The presence of thin, extended lensed radio arcs and the milli-arcsecond resolution of the observation provide direct sensitivity to the presence of FDM granules in the halo of the lens galaxy. In this Letter, using a simple modelling procedure with conservative assumptions and no dependence on baryonic physics models, we show that competitive constraints on $\mdm$ can be inferred using this single observation.

\section{Method} \label{sec:method}

\subsection{Bayesian inference}

A radio interferometer measures visibilities, or Fourier modes of the sky surface brightness distribution. Hence, the data $\data$ are a vector of complex visibilities, and the instrumental response is a discrete Fourier transform operator $\dft$. We represent the source as a vector $\source$ of pixelated surface brightness values on an adaptive Delaunay grid (see \citealt{vegetti2008}). Light from the source plane is mapped onto the image plane by the lens operator $\lensop$, which depends on the surface mass distribution of the lens galaxy. We describe this lens mass model using a set of parameters $\etalens$ for the smooth lens model, together with a field of pixelated potential perturbations $\dpsi$. The complete forward model is
\begin{equation}\label{eq:fmodel}
\model= \dft \lensop(\dpsi,\etalens) \source.
\end{equation}

In the present work, we wish to infer a posterior distribution on the dark matter particle mass $\mdm$, given the data $\data$. In equation \eqref{eq:fmodel}, $\etalens$ and $\source$ are nuisance parameters over which we marginalize during the inference process.  $\dpsi(\mdm,\fdm,\sigv)$ is the perturbation to the lensing potential due to the presence of FDM granules (fluctuations in the projected surface mass density), which is dependent on $\mdm$, $\fdm$ (the projected dark matter fraction within the Einstein radius of the lens), and $\sigv$ (the velocity dispersion of the dark matter in the lens); we detail the process for generating $\dpsi$ in Section \ref{sec:granules}. We also treat $\fdm$ and $\sigv$ as nuisance parameters.

We now turn to the inference on $\mdm$, which is encoded in $\dpsi$. We obtain a sample likelihood $P(\data \mid  \dpsi, \etalens,\lams)$ with respect to the lens model parameters ($\dpsi$, $\etalens$) and the source regularization weight $\lams$ using a linear source inversion (\citealt{powell2021,powell2022}; see also \citealt{suyu2006,vegetti2008,rybak2015b, hezaveh2016b,rizzo2018}), which simultaneously marginalizes over $\source$.  However, a key aspect of our modelling procedure is that $\dpsi$ is not deterministic with respect to $\mdm$, $\fdm$, or $\sigv$. Rather, any given ($\mdm$, $\fdm$, $\sigv$) can describe an infinite number of possible configurations of the FDM potential $\dpsi$. To make this clear in our notation, we label an individual likelihood as  $P_i(\data \mid  \mdm,\fdm,\sigv, \etalens,\lams)$,
indicating that $P_i$ is the sample likelihood obtained using the $i^\mathrm{th}$ possible realization of $\dpsi(\mdm,\fdm,\sigv)$, given fixed  $\etalens$ and $\lams$. The ordering of $i$ is arbitrary in practice, as the $\dpsi$ are generated randomly (Section \ref{sec:granules}).

With the sample likelihoods $P_i$ in hand, we phrase our inference on $\mdm$ as follows.  To accommodate the stochasticity of $P_i$, we build an empirical posterior $\post(\mdm)$ for each mass bin $[\mdm,\mdm+\,\Delta\mdm]$ by accepting or rejecting samples based on the likelihood ratio $P_i / \pthresh$. We define $\Delta \log P_i \equiv \log P_i - \log \pthresh$, accepting only FDM lens realizations with $\Delta \log P_i > 0$. $\pthresh$ is a threshold determined by the fiducial smooth lens model (see Section \ref{sec:smoothmodel}); we set $\pthresh$ at a conservative value corresponding to the $3\sigma$ contours in the posterior distribution of the smooth lens model parameters.  The remaining nuisance parameters $(\fdm,\sigv,\etalens, \lams)$ are naturally marginalized out during this sampling procedure.  $\post(\mdm)$ is therefore the probability that a realization of an FDM lens with the given $\mdm$ explains the data at least as well as the worst $0.3$ per cent of smooth models. This acceptance criterion is intentionally conservative, but we will see in Section \ref{sec:results} that the resulting constraint on $\mdm$ is still quite stringent.

\subsection{Smooth lens model} \label{sec:smoothmodel}

\begin{table} 
\caption{ Summary of parameters and priors. See Sections \ref{sec:smoothmodel} and \ref{sec:granules} for detailed descriptions and motivations for our prior choices.}
\centering
\begin{tabular}{l l l}
\hline
\noalign{\vskip 0.1cm}
Parameter & Description & Prior  \\
\noalign{\vskip 0.05cm}
\hline
\noalign{\vskip 0.1cm}
$\log_{10}(\mdm)$ & DM particle mass (eV) & $\mathcal{U}(-21.5,-19.0)$  \\
$\fdm$ & Projected DM mass fraction & $\mathcal{U}(0.5, 0.8)$  \\
$\sigv$ & DM velocity dispersion (km/s)  & $\mathcal{U}(100, 110)$  \\
$\etalens$ & Smooth lens model parameters & \multirow{2}{*}{$\mathcal{N}(\prmean,\prcov)$} \\
$\lams$ & Source regularization strength &  \\
\noalign{\vskip 0.05cm}
\hline
\end{tabular}
\label{tab:priors}
\end{table}

We model the smooth component of the lens using a composite model consisting of an elliptical power-law profile for the dark matter, a \sersic profile for the baryonic mass, multipole perturbations capturing internal angular complexity in the lens galaxy, and a third-order Taylor expansion of the external potential that corresponds to tidal effects from nearby field galaxies. This is the best composite lens model obtained for this system by \citet{powell2022}, where it is labeled PL+MP+SR+EP. We draw smooth lens model realizations $\etalens$ and source regularization strengths $\lams$ from a joint Gaussian prior that was fit to the posterior distribution on the PL+MP+SR+EP parameters from \citet{powell2022}; see Table \ref{tab:priors}.  

\begin{figure*}
 \centering
   \includegraphics[scale=\figscale]{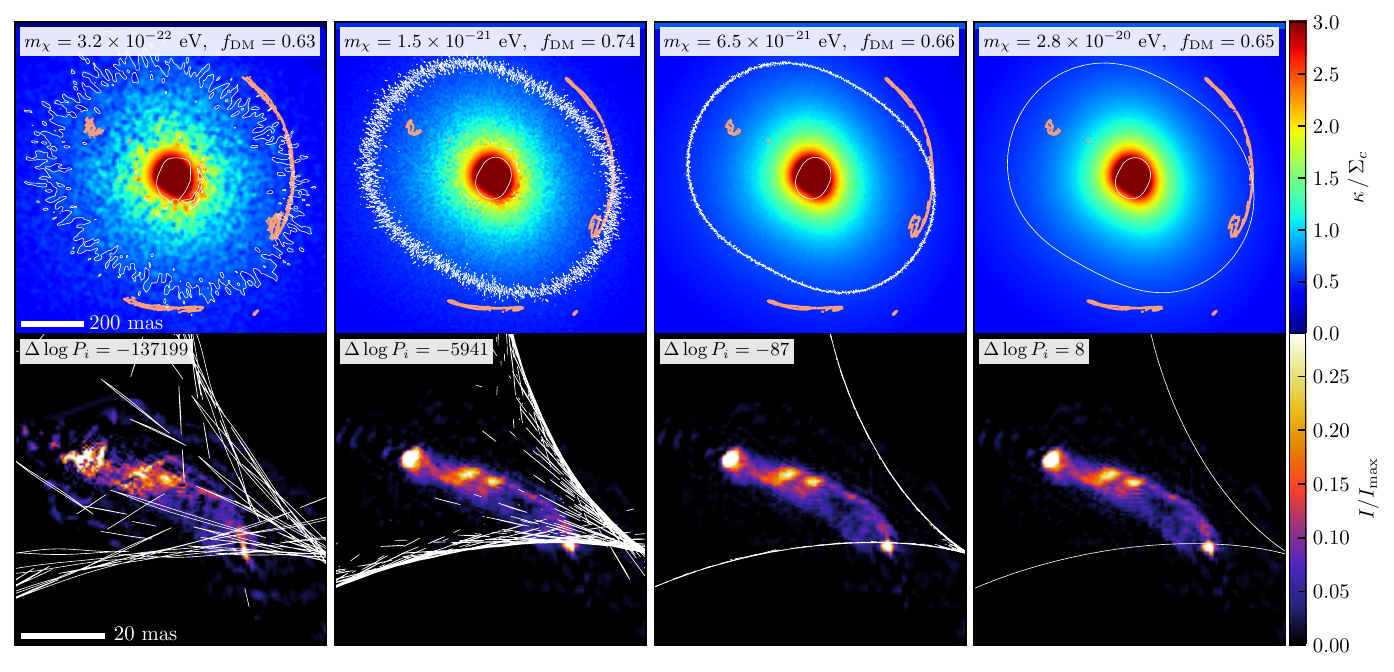}
  \caption{Example surface mass density maps ($\kappa$, in units of the critical density $\Sigma_c$) with the model lensed images in orange contours (top row) and the corresponding reconstructed source surface brightness maps ($I$, in units of the peak surface brightness $I_\mathrm{max}$; bottom row) for three random realizations of \jzsfo in an FDM cosmology. Critical curves and caustics are plotted in white. The lensing effect of the FDM granules is apparent: The critical curves wiggle back and forth across the lensed arcs, which would require the presence of multiple images of the same region of the source along the arc.  In the absence of such features in the observed data, the morphology of the inferred source is disrupted as the model attempts to fit the observation.}
  \label{fig:convergence}
\end{figure*}

The smooth model on its own provides a very accurate fit to the observed data.  The internal multipoles and external potential expansion capture deviations from perfect ellipticity on scales $\gtrsim100 \units{mas}$,  which yields an extremely clear and well-focused model image of the radio jet in the source plane \citep{powell2022}.  To obtain an FDM version of this lens, we simply perturb the smooth lens model with density granules (Section \ref{sec:granules}), as the mean density profile of an FDM halo outside of the central core is expected to be consistent with CDM \citep{hu2000,marsh2015}.  Small-scale perturbations to this lens model have an easily discernible effect on the inferred source morphology and hence the likelihoods $P_i$ (see Fig. \ref{fig:convergence}), which is the basis for our inference.

\subsection{Fuzzy dark matter granules} \label{sec:granules}

We generate random realizations of FDM granules using the model described by \citet{chan2020}. They derive expressions for the statistics of spatially-varying surface mass density fluctuations, given the density profile of the dark matter, as well as some basic assumptions on the behavior of scalar fields in a potential well.  In this model, the perturbation, $\delta \kappa$, in the lensing convergence (the projected surface mass density) due to the presence of FDM granules takes the form of a Gaussian random field with correlation length $\ldb$ and a position-dependent variance given by
\begin{equation}
    \label{eq:dk}
    \langle \delta \kappa ^ 2 \rangle = \frac{\ldb \sqrt{\pi}}{\Sigma_c^2} \int \rho_\mathrm{DM}^2 \, dl,
\end{equation}
where the integral is along the line of sight, $\rho_\mathrm{DM}$ is the smooth 3D density profile of the dark matter component of the lens, $\Sigma_c$ is the lensing critical surface mass density, and $\ldb=\hbar/(\mdm \sigv)$ corresponds to the (reduced) de Broglie wavelength of the dark matter particle.  In practice, we generate realizations of $\delta \kappa$ by first generating a white noise field modulated by the variance in equation \eqref{eq:dk}, then correlating using a Gaussian kernel of width $\ldb$ via an FFT-based convolution. We then solve for the resulting perturbation to the lensing potential $\dpsi$ using another FFT.

The correlation length $\ldb$ is inversely proportional to $\sigv$, the velocity dispersion of the dark matter in the lens galaxy, which is a proxy for the depth of the gravitational potential well in which the dark matter field resides. There are no resolved kinematic data on this lens system, so it must be estimated using the Einstein radius of the lens. \citet{alloin2007} found $\sigma_v = 101\units{km\,s^{-1}}$, using a cored pseudo-isothermal density profile. We derive $\sigma_v = 108\units{km\,s^{-1}}$, assuming a singular isothermal profile. To accommodate this uncertainty, we draw $\sigv$ from a uniform prior between 100 and 110 $\units{km\,s^{-1}}$ (see Table \ref{tab:priors}).

An additional source of uncertainty in generating FDM lens realizations is the dark matter fraction in the lens, $\fdm$, which directly determines the granule amplitude. Our composite smooth model from \citet{powell2022} gives a baryonic mass (measured within the critical curve) of $8.6\times10^{9}\units{\msun}$.  This number is in good agreement with observations by the \textit{Hubble Space Telescope} (HST) WFPC2 as part of the CfA-Arizona Space Telescope LEns Survey (CASTLES) project (e.g. \citealt{castles2000}); a fit to the $V$- and $I$-band lens galaxy photometry using \software{kcorrect} \citep{blanton2007} yields a baryonic mass of $8.0\times10^9\units{\msun}$.  The total projected mass of the lens within the critical curve is set by the Einstein radius at $2.7\times10^{10}\units{\msun}$.  Allowing for an uncertainty of $\pm0.2\units{dex}$ in the baryonic mass, we adopt a uniform prior on $\fdm$ between 0.5 and 0.8 (see Table \ref{tab:priors}). This prior range is consistent with dark matter fractions in massive early-type lens galaxies studied by \citet{oldham2018}.

We assume that all small-scale inhomogeneities in the lensing convergence are produced by FDM granules in the lens itself. We do not explicitly consider the effects of a central soliton core in the FDM halo; such a core would be much smaller than the Einstein radius of the lens \citep{schive2014b,chan2020}, and would therefore be absorbed in the smooth lens model.  Unlike the analysis by \citet{laroche2022}, we do not include subhalo or line-of-sight (LOS) halo populations in our lens model. This choice is justified because in the mass range of $\mdm \sim 10^{-22}$ to $10^{-20}\units{eV}$, in which our analysis is most sensitive, an FDM cosmology cannot produce subhaloes or LOS haloes that are highly concentrated or numerous enough to mimic the signal of FDM granules (\citealt{schive2016}; see also Fig. 5 of \citealt{laroche2022}); indeed, any large-scale contribution to the lens model by diffuse low-mass haloes would already be accounted for in the smooth model.  The practical effects of excluding low-mass haloes from our model are the loss of some sensitivity to $\mdm$ and the inability to place an upper bound on $\mdm$.

\section{Results} \label{sec:results}

 \begin{figure*}
 \centering
   \hspace*{-7.75pt}%
   \includegraphics[scale=\figscale]{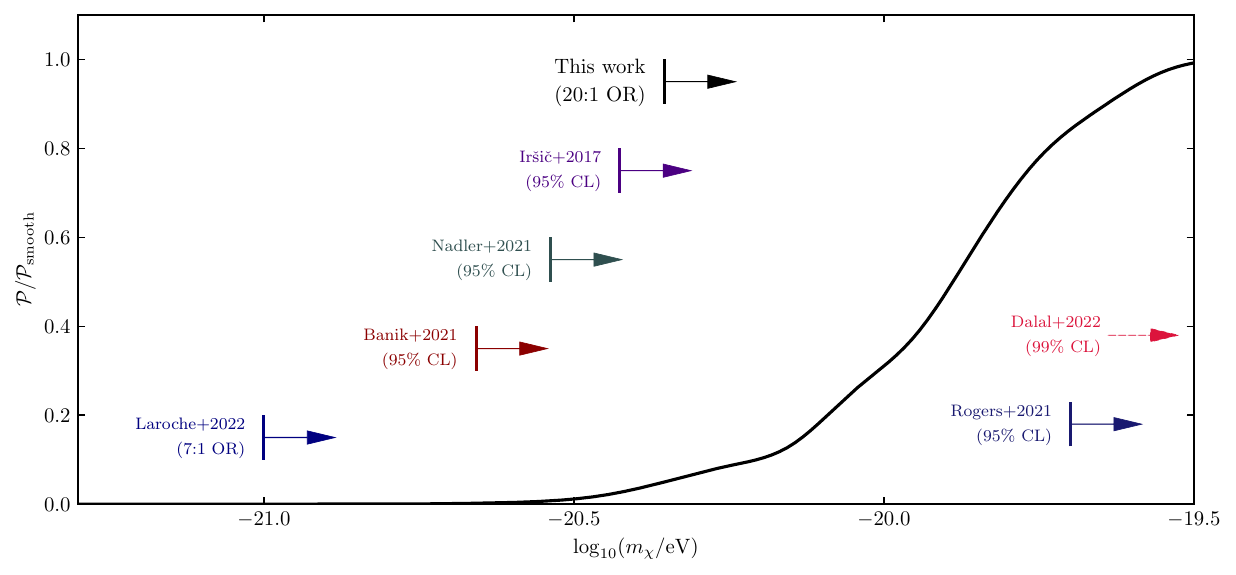}
  \caption{Summary of the main result in this work and comparison to other lower bounds on $\mdm$. The fully marginalized posterior odds ratio (POR) as a function of $\mdm$ is plotted as a solid black curve. We superimpose lower bounds on $\mdm$ for this work (where $\mdm \leq 4.4\times10^{-21}\units{eV}$ is excluded with a 20:1 POR) alongside several other observational constraints (see Section \ref{sec:intro}).  The lower bound of $\mdm > 3\times10^{-19}\units{eV}$ by \protect\citet{dalal2022} lies beyond the plot axis. We give the strength of each constraint as presented in the original work, either as a confidence level (CL) or as an odds ratio (OR; either posterior or likelihood). The vertical positions and colours of the lower bounds on this plot are aesthetic and have no physical meaning.} 
  \label{fig:likelihood}
\end{figure*}

We show example convergence maps for three FDM lens realizations with their corresponding maximum a-posteriori (MAP) source surface brightness reconstructions in Fig. \ref{fig:convergence}.  For $\mdm \lesssim 10^{-21}\units{eV}$, the critical curves (plotted in white) cross back and forth many times across the lensed arcs. Such a configuration of critical curves would imply the presence of many images of alternating parity along the arc from the same region of the source, where the corresponding caustics (lower panels, plotted in white) zig-zag across it. This phenomenon was first pointed out by \citet{chan2020}, who predict extra images of point-like lensed quasars in FDM cosmologies. When we impose a potential perturbation $\dpsi$ in the lens model, but no compatible extra image features are present in the data, then the forward model has no choice but to attempt to fit the observed data using a highly disrupted source surface brightness distribution. (In the case of a parametric source model, this disruption would instead appear in the model residuals). Arcs that lie away from the critical curves, though not containing extra images, also impart some disruption to the inferred source morphology.  In the bottom row of Fig. \ref{fig:convergence}, we observe the presence of spurious discontinuities and misalignments of the back-projected source components. These features are penalized in the sample likelihoods $\Delta\log P_i$, whose values are inset in the bottom row for each source reconstruction.

To construct the posterior $\post(\mdm)$, we compute likelihoods for $4.1\times10^4$ sample FDM lens realizations with $\mdm$ drawn from the log-uniform prior range $\log(\mdm/\mathrm{eV}) \in [-21.5, -19.0]$. Of these, $\sim37$ per cent meet the acceptance criterion $\Delta\log P_i > 0$.  We collect the accepted samples in bins of width 0.1 dex to arrive at the posterior $\post(\mdm)$.  We present the resulting constraint on $\mdm$ in terms of the \emph{posterior odds ratio} (POR) between FDM with a particle mass $\mdm$ and the smooth model, $\post/\psmooth$. Since we have defined the sample acceptance threshold $\pthresh$ relative to the $3\sigma$ contours in the smooth model posterior, this is equivalent to a simple rescaling of $\post$ by a factor of 0.997.  We plot $\post/\psmooth$ as a function of $\mdm$ in Fig. \ref{fig:likelihood}.  We find $\mdm \leq 4.4\times10^{-21}\units{eV}$ to be disfavored relative to the smooth model with a 20:1 POR, which we interpret as a lower bound of $\mdm > 4.4\times10^{-21}\units{eV}$.  We also see that $\post/\psmooth \rightarrow 1$ at $\mdm = 3.2 \times 10^{-20} \units{eV}$, meaning that at this particle mass our analysis cannot distinguish between an FDM lens and the fiducial smooth model.

\section{Discussion and conclusions} \label{sec:discussion}

Our results are consistent with other observational lower bounds on $\mdm \gtrsim 2\times10^{-21}\units{eV}$, particularly those obtained via modelling of the substructure populations in the Milky Way \citep{nadler2021a, banik2021} and around galaxy-scale strong gravitational lens systems \citep{laroche2022}. However, this work instead relies exclusively on the strongest and theoretically best-understood phenomenological prediction of FDM, which is the formation of granules in the main halo.  Even under simplifying conditions in which we ignore the presence of subhaloes (which in the $\mdm$ range considered here are diffuse enough to be accounted for \emph{a priori} by the smooth lens model), and a very generous sample acceptance threshold relative to the smooth model (Section \ref{sec:method}), we are able to rule out FDM models solely by considering the perturbative effect of these granules on extended gravitationally lensed arcs observed at milli-arcsecond angular resolution. 

Our results generalize in a simple way to vector ($s=1$) and higher-spin ($s>1$) boson fields \citep{Amin:2022pzv}, or equivalently, FDM composed of $N$ multiple fields of equal $\mdm$ \citep{Gosenca:2023yjc}.  The presence of $N=2s+1$ degrees of freedom in the FDM attenuates the granule amplitude by a factor of $1/\sqrt{N}$, which translates to a rescaling of the particle mass $\mdmc{N} = \mdm/N$ (see \citealt{Amin:2022pzv}), where $\mdm$ is the single scalar field result derived in this work. Hence, for vector boson DM with $s=1$ ($N=3$), we obtain a lower bound of $\mdmc{\mathrm{vec}} > 1.4 \times 10^{-21}\units{eV}$.

While this work and that of \citet{laroche2022} both infer $\mdm$ directly from strong gravitational lens observations, there are key differences that endow our analysis with the sensitivity to constrain $\mdm$ using a single observation. First and foremost is the angular resolution of the observation. While this VLBI observation has an effective point spread function (PSF) width of $5.5 \times 1.8 \units{mas}^2$, the 11 observations used by \citet{laroche2022} were observed using either adaptive optics on the W.M. Keck Observatory or the WFC3 on the HST with an angular resolution no better than $\sim 70\units{mas}$, giving information only on the relative positions and fluxes of the unresolved quasar images. By contrast, the VLBI observation of the resolved long, thin arcs in \jzsfo is sensitive to the source morphology itself. Indeed, a simple calculation of the value of $\mdm$ corresponding to a projected (reduced) de Broglie wavelength of $\ldb = 1.8\units{mas}$ (the minor axis of the PSF) yields $\mdm \sim 2\times10^{-21}\units{eV}$, suggesting that our sensitivity to $\mdm$ is limited by angular resolution.

The modelling procedure for unresolved quasar images by \citet{laroche2022} requires special care, as the data are not as informative. This includes the selection of lens systems that do not contain stellar disks (which can masquerade as dark-matter-induced flux ratio anomalies; \citealt{gilman2017,hsueh2018b,he2022}), considerations of source compactness and variability \citep{hsueh2020}, as well as the inclusion of an explicit model for subhaloes and LOS haloes. The latter is especially important, as FDM granules and low-mass haloes produce the same observable effect on an unresolved compact image. 

Our results demonstrate that with the milli-arcsecond angular resolution of VLBI, competitive constraints on dark matter models can be inferred using a single strong gravitational lens observation.  They also demonstrate for the first time the use of high-resolution observations to directly search for FDM granule structures.  The constraints presented here can be improved primarily by increasing the angular resolution; for example, follow-up of known radio lenses with global VLBI in the $10$ to $100\units{GHz}$ range would extend our sensitivity to $\mdm$ by one to two orders of magnitude in mass.  Source structure and lensing configuration are also very important; given a fixed resolution, an extremely bright and compact source lying exactly on a lensing caustic would be stretched into long, smooth arcs exhibiting much less structure than \jzsfo. Any perturbation to these arcs would unambiguously  be a gravitational perturbation by low-mass structures in the lens galaxy or along the LOS. While this type of source and lensing configuration would be ideal for inferring dark matter constraints, at the moment only a handful of such systems are known to exist. However, in coming years the \textit{Euclid} mission and the Square Kilometre Array (SKA) will discover many thousands of new galaxy-scale strong lenses. We therefore expect the number of known strong lens systems that are useful for this type of analysis to increase by orders of magnitude; high-resolution follow-up observations of these systems will position VLBI as a leading observational tool for constraining the particle nature of dark matter.

\section*{Data availablity}

The interferometric data used in this Letter are available via the JIVE archive. HST WFPC2 data from the CASTLES Survey are available at \nolinkurl{https://lweb.cfa.harvard.edu/castles/}. 

\section*{Acknowledgements}

We thank Mustafa Amin, Chris Fassnacht, and Jowett Chan for helpful input. DP and SV acknowledge funding from the European Research Council (ERC) under the European Union's Horizon 2020 research and innovation programme (LEDA: grant agreement No 758853). 
SM acknowledges support by the National Science Foundation under Grant No.\ 2108931. 
JPM acknowledges support from the Netherlands Organization for Scientific Research (NWO, project number 629.001.023) and the Chinese Academy of Sciences (CAS, project number 114A11KYSB20170054).
The National Radio Astronomy Observatory is a facility of the National Science Foundation operated under cooperative agreement by Associated Universities, Inc. The European VLBI Network is a joint facility of European, Chinese, South African and other radio astronomy institutes funded by their national research councils. Scientific results from data presented in this publication are derived from the following EVN project code: GM070.


\bibliographystyle{mnras}
\bibliography{references} 






\bsp	
\label{lastpage}
\end{document}